\documentclass[english,11pt,nofootinbib,showpacs,notitlepage]{revtex4-1}
\usepackage[T1]{fontenc}
\usepackage{babel}
\usepackage[utf8]{inputenc}
\usepackage{amssymb,mathrsfs,amsfonts,amsmath,euscript,textcomp,graphicx,wrapfig,color,txfonts,lmodern,multirow,bigdelim,dsfont}
\definecolor{darkgreen}{rgb}{0,.7,0}
\usepackage[colorlinks,linkcolor=blue,citecolor=darkgreen,pagebackref=true]{hyperref}
\def\[{\left[}
\def\]{\right]}
\def\({\left(}
\def\){\right)}

\def\1{{\bf CI}}
\def\2{{\bf CII}}
\def\3{{\bf CIII}}

\def\m{\mathcal{M}}

\newcommand{\eq}[1]{\begin{equation}#1\end{equation}}

\newcommand{\const}{\mathop{\rm const}\nolimits}

\newcommand{\bgsb}[1]{\bigl[#1\bigr]}

\newcommand{\lrp}[1]{\left(#1\right)}

\newcommand{\nd}{\noindent}

\newenvironment{tightcenter}{%
  \setlength\topsep{0pt}
  \setlength\parskip{0pt}
  \begin{center}
}{%
  \end{center}
}

\begin{document}

\title{Dynamic compactification with stabilized extra dimensions in cubic Lovelock gravity}

\author{Dmitry Chirkov}
\affiliation{Bauman Moscow State Technical University, Moscow, Russia}
\author{Alex Giacomini}
\affiliation{Instituto de Ciencias F\'isicas y Matem\'aticas, Universidad Austral de Chile, Valdivia, Chile}
\author{Alexey Toporensky}
\affiliation{Sternberg Astronomical Institute, Moscow State University, Moscow, Russia}
\affiliation{Faculty of Physics, Higher School of Economics, Moscow, Russia}

\begin{abstract}
In this paper  the dynamic compactification in Lovelock gravity with a cubic term is studied. The ansatz will be of space-time where the three dimensional space and the extra dimensions are constant curvature manifolds with independent scale factors.
The numerical analysis shows that there exist a phenomenologically realistic compactification regime  where the three dimensional hubble parameter and the extra dimensional scale factor tend to a constant. This result comes as surprise as in Einstein-Gauss-Bonnet gravity this regime exists only when the couplings of the theory are such that the theory does not admit a maximally symmetric solution (i.e. "geometric frustration"). In cubic Lovelock gravity however there always exists at least one maximally symmetric solution which makes it fundamentally different from the Einstein-Gauss-Bonnet case.
Moreover, in opposition to Einstein-Gauss-Bonnet Gravity, it is also found that for some values of the couplings and initial conditions these compactification regimes can coexist with isotropizing solutions.

\end{abstract}


\maketitle

\section{Introduction}
A feature which makes gravity unique among all other fundamental interactions is that it is described by space-time geometry. As space-time becomes a dynamical object it is, in principle, possible that it may  have more than four dimensions. This hypothesis is also encouraged by string theory which is consistent only in higher dimensional space-times.
Moreover, since the original idea of Kaluza and Klein \cite{KK1, KK2, KK3}, the existence of extra dimensions can be used  to obtain the fundamental gauge fields from pure geometry. As these extra dimensions, at least macroscopically, can not be observed it is reasonable to suppose that they are compactified to a very small scale.

In order to implement this idea it is necessary to find a reasonable explanation why  space-time should prefer to compactify instead of having all space dimensions of similar size. Especially it may be that in the far past the extra dimensions were of similar size than the three dimensional part and only at a later stage the universe compactified. It is therefore necessary to make an analysis of the dynamical evolution of the three dimensional and extra dimensional part of space. This opens the question which gravity theory to consider. A natural guess would be just to use the higher dimensional Einstein-Hilbert action (plus Lambda term). This idea is attractive due to the simplicity of the principles
from which it is built namely to be constructed with curvature invariants and that it leads to second order derivative field equations in the metric.
Indeed one can proof that in four dimensional space time the E-H action (plus lambda term) is the only action that can be built from these basic principles.
To be  precise in four dimensions it is possible to add to the E-H action a so called Gauss-Bonnet term $R_{\mu \nu \alpha \beta}R^{\mu \nu \alpha \beta}-4R_{\mu \nu}R^{\mu \nu}+ R^2$ but which does not affect the equations of motion being an Euler density. However in dimensions  higher than four the GB term does affect the equations of motions but this correction remains of second order in the derivatives of the metric. This means that there is no good reason to discard a GB term in the higher dimensional gravity action  as it respects the same principles according to which one builds the four dimensional EH action. It is also worth to point out that in some string theory the effective low energy limit is described by EGB gravity rather than General Relativity \cite{GastGarr}. It turns out that EGB gravity is just a special case of a more general gravity theory called Lovelock Gravity \cite{Lovelock}. Indeed increasing the space-time dimensions in every new odd dimension it is possible to add a higher curvature power term to the action whose correction to the equations of motion again are only of second order in the derivatives of the metric. These  Lovelock terms are the dimensional continuations of Euler densities of the lower even dimensional space-time. In arbitrary space-time dimensions  Lovelock Gravity is therefore the most natural extension of General Relativity to higher dimensions. An interesting feature of Lovelock gravity, which is absent in GR, that in the first order formalism the equations of motion do not imply the vanishing of torsion \cite{TZ-CQG}. Some examples of exact solutions with non trivial torsion have been found in \cite{CGT07, CGW07, CG, CG2, ACGO}.  Here however we will consider only the zero torsion sector.

Exact solutions describing static compactified  space-times which are a direct product of a four dimensional Lorentzian manifold times a Euclidean extra dimensional space are known in literature as spontaneous compactification. In higher dimensional GR with lambda spontaneous compactification only exist when the curvature scale of the extra dimensions is of the same magnitude as the one of the four dimensional space-time.  Spontaneous compactification can be achieved in EGB gravity but as problem arises that in the four dimensional part of space-time  General Relativity is not recovered as an effective theory as the equations of motion impose an additional scalar constraint on the four dimensional Euler density \cite{GOT}. Spontaneous compactification  in this case has been studied \cite{add_1}. For cubic Lovelock theory it is possible, for some values of the couplings, to get spontaneous compactification which recovers GR in four dimensions and with arbitrarily small extra dimensions \cite{CGTW09}.

In the context of cosmology it is of course necessary to study for the time evolution of the size of the three dimensional space and the extra dimensions. Dynamical compactification with time dependent scale factors has been studied for various models in \cite{DFB, MH, DCMTP, MM}. In the context of cosmology the most studied Lovelock  Gravity is EGB gravity. In $5+1$  dimensional EGB gravity  the dynamical compactification has been studied in \cite{EMOOF}. Dynamical compactification in EGB gravity with several ansatz for the scale factors has been studied in \cite{Maeda1, Maeda2}.

Speaking of compactification, it is worth to mention a quite viable compactification models basing on exact exponential solutions. Solutions with exponential time dependence of the scale factor in Gauss-Bonnet gravity (without the Einstein term in the action) has been found in~\cite{Ivashchuk-1} and then have been generalized to full EGB theory in~\cite{grg10, KPT}. A study of such solutions in EGB gravity has revealed an interesting fact that
they exists only if the space has isotropic subspaces~\cite{CST1}; this fact remains valid also for general Lovelock model~\cite{ChPavTop1}. It should be emphasized that this division is not introduced "by hand" and appears naturally from equations of motion as a condition for such solutions
to exist. Moreover, it appears that there are solutions where three dimensions (corresponding to "our real world") are expanding, while remaining are contracting making the compactification viable. Stability of such solutions have been studied in~\cite{PavlStab,Ivashchuk-2,Ivashchuk-3,ChTop,Ivashchuk-4}. As for power-law
solutions in general Lovelock gravity, they have been studied in \cite{PavlGen, Dadhich}.

There is however a good reason to study the effect of the addition of a cubic term to the gravitational action.
This is due to the fact  that Lovelock theories can be divided in two  subclasses according to if the highest term in the action is an even or odd power in the curvature.
The reason for this is the following: In General Relativity the Lambda term (zero order in the curvature) in the EH action gives the curvature scale of its maximally symmetric solution (de Sitter, Minkowski or Anti de Sitter depending if the Lambda term is positive zero or negative). If one adds higher power Lovelock terms to the action the zero Lovelock term is no more directly proportional to the curvature scale of the maximally symmetric solution. Indeed if one plugs the ansatz of a maximally symmetric space-time in the Lovelock equations of motion where the highest term is of n-th order in the curvature one gets an n-th degree polynomial equation in the curvature scale of the maximally symmetric space-time. This especially means that, depending on the exact  value of all the Lovelock couplings one can have up to $n$ maximally symmetric solutions. The value of the curvature of these solutions depend on all Lovelock couplings and just on the zero term like in General relativity. If the highest term of the Lovelock action is an even power in the curvature there exist an open region in the coupling constants space where the polynomial has no real roots and therefore there does not exist maximally symmetric solution at all. This situation is also known as "geometric frustration". It was shown that in EGB cosmology it is possible to achieve a phenomenologically realistic dynamical compactification scenario (i.e. asymptotic constant three dimensional hubble parameter and scale factor of extra dimensions shrinking to a constant non-zero value) only in the case of geometric frustration \cite{Canfora-Giacomini-Pavluchenko-1, Canfora-Giacomini-Pavluchenko-2}. Adding matter to the action it is also possible in this scenario to recover the Friedmann regime \cite{CGPT}.
In the case that the highest term of the Lovelock action is odd in the curvature the polynomial defining the curvature of the maximally symmetric solutions has always at least
one real root. This means that in this case geometric frustration  does not occur.
This fact suggests that Lovelock theories with an odd power of the curvature as highest term have a fundamentally different cosmological behavior than theories with even highest power. Therefore, in order to study cosmological dynamics, it is great theoretical interest to study the effect of an odd power curvature term the simplest one being of course cubic. The cubic term exist when the total dimension of space-time is at least seven.
Due to the absence of geometric frustration the addition of a cubic Lovelock term to the gravitational action can potentially bear the risk that  it is not possible to achieve compactification with stabilized extra dimensions.

However in this paper we will show that the addition of cubic term to the gravitational action remarkably does not spoil the existence of such a compactification regime. A new feature in the cubic theory is also the coexistence of compactification regime with isotropization of space-time. In contrast in EGB cosmology isotropization can coexist only with extra dimensions exponentially shrinking to zero radius \cite{PT}.
In order to perform the analysis of dynamical compactification we will use an ansatz of a space-time of the form
$ds^2=-dt^2+a(t)^2d\Sigma^2_{3}+b(t)^2d\Sigma^2_{D}$ where the manifolds $\Sigma_3$ and $\Sigma_D$ are constant curvature and represent the three dimensional space and the extra dimensions respectively. For a physically realistic compactification model the scale factor of the extra dimensions should shrink to a constant nonzero value and the hubble parameter of the three dimensional space should tend to a constant. Remarkably the numerical analysis shows that for certain values of the couplings and initial values
there exist a coexistence of compactification and isotropization regimes.

The structure of the paper is the following: In the next section the equations of motion are derived. In the third section the numerical analysis is performed and in the last section the conclusions will be given.

\section{Equations of motion\label{setup}}
\nd The Lovelock action in arbitrary dimensions in the vielbein formalism reads
\eq{\int\varepsilon_{A_1\ldots A_{D+4}}\sum\limits_{k=0}^{\lfloor\frac{D+4}{2}\rfloor}\frac{c_k}{D+4-2k}\bigwedge\limits_{n=1}^{k}R^{A_{2n-1}A_{2n}}
\;\wedge\;\bigwedge\limits_{m=2k+1}^{D+4}e^{A_m}}
where $\varepsilon_{A_1\ldots A_{D+4}}$ are the Levi-Civita symbols, $R^{A_{2n-1}A_{2n}}$ are the Riemannian curvature forms, $e^{A_1},\ldots,e^{A_{D+4}}$ is the vielbein basis, $c_k$ are coupling constants. Varying it with respect to the vielbein we obtain the equations of motion:
\eq{E_{A_1}\equiv\varepsilon_{A_1\ldots A_{D+4}}\sum\limits_{k=0}^{\lfloor\frac{D+3}{2}\rfloor}c_k\cdot\bigwedge\limits_{n=1}^{k}R^{A_{2n-1}A_{2n}}
\;\wedge\;\bigwedge\limits_{m=2k+1}^{D+4}e^{A_m}=0\label{EoM-initial}}
We will make an ansatz of a warped product space-time of the form $\m_4\times \m_D$ where $\m_4$ is a Friedman-Robertson-Walker
manifold with scale factor $a(t)$ whereas $\m_D$ is a $D$-dimensional Euclidean compact and constant curvature manifold with scale factor $b(t)$. The ansatz for the metric is
\eq{ds^2=-dt^2+a(t)^2d\Sigma^2_{3}+b(t)^2d\Sigma^2_{D}}
where $d\Sigma^2_{3}$ and $d\Sigma^2_{D}$ stand for the metrics of constant curvature manifolds $\Sigma_{3}$ and $\Sigma_{D}$ respectively. The Riemannian curvature forms read:
\eq{R^{0i}=\frac{\ddot{a}}{a}e^0\wedge e^i,\quad R^{0a}=\frac{\ddot{b}}{b}e^0\wedge e^a,\quad R^{ia}=\frac{\dot{a}}{a}\frac{\dot{b}}{b}e^i\wedge e^a,\quad R^{ij}=\frac{\gamma_{3}+\dot{a}^2}{a^2}e^i\wedge e^j,\quad R^{ab}=\frac{\gamma_{D}+\dot{b}^2}{b^2}e^a\wedge e^b\label{R}}
where $\gamma_{3}$ is constant Riemannian curvature of spatial submanifold of the manifold $\m_4$, $\gamma_{D}$ is constant Riemannian curvature of $\m_D$; here and after we will use Latin indices from the start of the alphabet for the manifold $\m_D$ (i.e. $a,b,c\ldots$ run from 4 to $D+4$) and Latin indices from the middle of the alphabet for spatial submanifold of the manifold $\m_4$ (i.e. $i,j,k\ldots$ run from 1 to 3). In what follows we assume that "our" (3+1)-dimensional world is flat ($\gamma_{3}=0$);  nonzero $\gamma_{3}$ does not affect the presence of the dynamical compactification regime, as discussed in~\cite{Canfora-Giacomini-Pavluchenko-1,Canfora-Giacomini-Pavluchenko-2}; the non-zero curvature for extra dimensions can be normalized as $\gamma_{D}=\pm1$. In what follows we consider the case $\gamma_D=-1$.


Since there are only two scale factors, one obtains two independent dynamical equations: first (we denote it by $E_1$) by varying the action with respect to any of $e^1,e^2,e^3$ vielbein elements, second (we denote it by $E_2$) by varying the action with respect to any of $e^4,\ldots,e^{D+4}$ vielbein elements;  we have also a constraint ($E_0$), which is obtained by varying the action with respect to $e^0$. Below we show in detail how one can derive these equations  for the particular case of  $D=7$.

Let us write down equation~(\ref{EoM-initial}) in an explicit way:
\eq{\begin{split}
      E_{A_1}=\varepsilon_{A_1\,A_2\,A_3\,A_4\,A_5\,A_6\,A_7}\Bigl[&c_0\cdot e^{A_2}\wedge e^{A_3}\wedge e^{A_4}\wedge e^{A_5}\wedge e^{A_6}\wedge e^{A_7}+c_1\cdot R^{A_2A_3}\wedge e^{A_4}\wedge e^{A_5}\wedge e^{A_6}\wedge e^{A_7}+ \\
        & +c_2\cdot R^{A_2A_3}\wedge R^{A_4A_5}\wedge e^{A_6}\wedge e^{A_7}+c_3\cdot R^{A_2A_3}\wedge R^{A_4A_5}\wedge R^{A_6A_7}\Bigr]
    \end{split}
}
Let $A_1=0$; splitting the rest of indices into the part corresponding to "our"\, 3D space ($i,j,k$) and part corresponding to extra dimensions ($A_{\beta}=a,b,c$), we get:
\eq{\begin{split}
      E_0=&\varepsilon_{0ijkabc}\Bigl[c_0\cdot C^3_6\cdot  e^{i}\wedge e^{j}\wedge e^{k}\wedge e^{a}\wedge e^{b}\wedge e^{c}+ \\
        & +c_1\bigl(2\cdot C^2_4\cdot R^{ia}\wedge e^{j}\wedge e^{k}\wedge e^{b}\wedge e^{c}+C^1_4\cdot R^{ij}\wedge e^{k}\wedge e^{a}\wedge e^{b}\wedge e^{c}+C^1_4\cdot R^{ab}\wedge e^{i}\wedge e^{j}\wedge e^{k}\wedge e^{c}\bigr)+ \\
        & +c_2\bigl(2\cdot 2\cdot R^{ij}\wedge R^{ab}\wedge e^{k}\wedge e^{c}+2\cdot 2\cdot R^{ij}\wedge R^{ka}\wedge e^{b}\wedge e^{c}+ \\
        & \hspace{5cm}+2\cdot 2\cdot 2\cdot R^{ia}\wedge R^{jb}\wedge e^{k}\wedge e^{c}+2\cdot 2\cdot R^{ab}\wedge R^{ci}\wedge e^{j}\wedge e^{k}\bigr)+ \\
        & +c_3\bigl(2\cdot 3!\cdot R^{ij}\wedge R^{ka}\wedge R^{bc}+2\cdot 2\cdot 2\cdot R^{ia}\wedge R^{jb}\wedge R^{bc}\bigr)
    \end{split}
\label{EoM-explicit}}
where $C^k_n=\frac{n!}{k!(n-k)!}$ is the number of ways to choose $k$ elements from a set of $n$ elements.

One can see that each term in~(\ref{EoM-explicit}) is multiplied by some factor. The reason for this lies in the fact that due to antisymmetry of  wedge product and the Levi-Civita symbols, interchanging spacial indices and extra-dimensional indices does not affect the result. For example, the sums $\varepsilon_{0ijkabc} R^{ia}\wedge e^{j}\wedge e^{k}\wedge e^{b}\wedge e^{c}$ and $\varepsilon_{0ijcabk} R^{ia}\wedge e^{j}\wedge e^{c}\wedge e^{b}\wedge e^{k}$ give the same result; so, the term $R^{ia}\wedge e^{j}\wedge e^{k}\wedge e^{b}\wedge e^{c}$ is multiplied by the factor $2\cdot C^2_4$ because there are $C^2_4$ ways to choose two spacial indices ($j,k$) from a set of four indices and, besides, interchanging indices $i$ and $a$ of the curvature form (simultaneously with interchanging corresponding indices of the Levi-Civita symbol) does not change the result -- that gives the factor 2. Analogously, in the sum $\varepsilon_{0ijkabc} R^{ij}\wedge R^{ka}\wedge e^{b}\wedge e^{c}$ one can interchange indices $k$ and $a$ of the curvature form as well as the forms $R^{ij}$ and $R^{ka}$ without changing the result, so the term $R^{ij}\wedge R^{ka}\wedge e^{b}\wedge e^{c}$ is multiplied by the factor $2\cdot 2$, etc.

Substituting~(\ref{R}) into~(\ref{EoM-explicit}), summing over the indices $i,j,k,a,b,c$ and replacing $D=7$ by a general $D$, we obtain:
\eq{E_T=\sum\limits_{k=0}^{\lfloor\frac{D+3}{2}\rfloor}
\sum\limits_{\alpha_{0a}=0}^{N_{0a}}\sum\limits_{\alpha_{0i}=0}^{N_{0i}}\sum\limits_{\alpha_{ij}=0}^{N_{ij}}\sum\limits_{\alpha_{ia=0}}^{N_{ia}}
c_k\cdot C^{T\,k}_{\alpha_{0a}\;\alpha_{0i}\;\alpha_{ij}\;\alpha_{ia}}\cdot
\lrp{\frac{\ddot{b}}{b}}^{\alpha_{0a}}\lrp{\frac{\ddot{a}}{a}}^{\alpha_{0i}}\lrp{\frac{\gamma_{3}+\dot{a}^2}{a^2}}^{\alpha_{ij}}
\lrp{\frac{\dot{a}\dot{b}}{ab}}^{\alpha_{ia}}\lrp{\frac{\gamma_{D}+\dot{b}^2}{b^2}}^{k-s},\label{ET}}
where $T=0,1,2$; $k$ is an order of Lovelock term; $\alpha_{0a},\alpha_{0i},\alpha_{ij},\alpha_{ia}$ are the numbers of the forms $R^{0a},R^{0i},R^{ij},R^{ia}$ respectively in a given term (see~(\ref{EoM-explicit})); $s=\alpha_{0a}+\alpha_{0i}+\alpha_{ij}+\alpha_{ia}$; $C^{T\,k}_{\alpha_{0a}\;\alpha_{0i}\;\alpha_{ij}\;\alpha_{ia}}$ generalizes factors in~(\ref{EoM-explicit}) and allows us take into account factors arising due to summation over the indices $i,j,k,a,b,c$:
\raggedbottom
\eq{\begin{split}
      &C^{T\,k}_{\alpha_{0a}\;\alpha_{0i}\;\alpha_{ij}\;\alpha_{ia}}=\frac{k!\,2^{\alpha_{0a}}\,2^{\alpha_{0i}}\,2^{\alpha_{ia}}}
{\alpha_{0a}!\,\alpha_{0i}!\,\alpha_{ij}!\,\alpha_{ia}!\,(k-\alpha_{0a}-\alpha_{0i}-\alpha_{ij}-\alpha_{ia})!}\cdot \vspace{.3cm}\\
      &\cdot\frac{(D+3-2k)!}{(1-\delta_{0T}-\alpha_{0a}-\alpha_{0i})!\,(3-\delta_{1T}-\alpha_{ia}-\alpha_{0i}-2\alpha_{ij})!\,
(D-1-2k+\delta_{0T}+\delta_{1T}+\alpha_{0a}+2\alpha_{0i}+\alpha_{ia}+2\alpha_{ij})!}
    \end{split}\label{C}}
$N_{0a},N_{0i},N_{ij},N_{ia}$ are the maximal numbers of the forms $R^{0a},R^{0i},R^{ij},R^{ia}$ respectively in a given term in~(\ref{ET}). These numbers are evaluated by the following formulas:
\eq{N_{0a}=H(k-1)(1-\delta_{0T}),\quad N_{0i}=H(k-1)(1-\delta_{0T}-\alpha_{0a})\label{0a,0i}}
\eq{N_{ij}=H(k-1)(1-\delta_{1k}(\alpha_{0a}+\alpha_{0i})-\alpha_{0i}\delta_{1T}H(k-2))\label{ij}}
\eq{\begin{split}
      N_{ia}=H(k-1)\Bigl\{&\delta_{1k}\bigl[1-\alpha_{0a}-\alpha_{0i}-\alpha_{ij}\bigr]+\\
        +&\delta_{2k}\bigl[2-\alpha_{0a}-\alpha_{0i}-\alpha_{ij}(1+\delta_{1T}H(-\alpha_{0a}))\bigr]+ \\
        +&\delta_{3k}\bigl[3-\delta_{1T}-(\alpha_{0a}+\alpha_{ij}(1+H(-\alpha_{0a})))(1-\delta_{1T})-\alpha_{0i}-2\alpha_{ij}\delta_{1T}\bigr]+\\
        +&H(k-4)\bigl[3-\delta_{1T}-\alpha_{0i}-2\alpha_{ij}\bigr]\Bigr\}
    \end{split}\label{ia}}
where
\eq{H(x)=\left\{\begin{array}{c}
                  0,\quad x<0 \\
                  1,\quad x\geqslant 0
                \end{array}\right.
                }
Now we explain how we have obtained equations~(\ref{0a,0i})-(\ref{ia}). First of all, for $k=0$ there are no any forms in the formula~(\ref{ET}) -- this is why we have the factor $H(k-1)$ in eqs.~(\ref{0a,0i})-(\ref{ia}): $H(k-1)=0$ for $k=0$ and $H(k-1)=1$ for $k\geqslant1$.

Let us consider the formulas~(\ref{0a,0i}). Equation $E_0$ ($T=0$) does not contain $R^{0a}$ and $R^{0i}$ forms since we obtain it by varying action with respect to the $e^0$ vielbein element, so we use the Kronecker delta $\delta_{0T}$ in the formulas for $N_{0a}$ and $N_{0i}$: if $T=0$ then $\delta_{0T}=1$ and both $N_{0a}$ and $N_{0i}$ equals to zero. Due to antisymmetry there should not be repeated indices in~(\ref{ET}), so if we have the form $R^{0a}\;(\alpha_{0a}=1)$, we can not have the form $R^{0i}$ at the same time and vice versa, therefore we subtract $\alpha_{0a}$ in the formula for $N_{0i}$ in~(\ref{0a,0i}).

Now we consider formula~(\ref{ij}). Since the equation $E_1$ ($T=1$) is obtained by varying the action with respect to one of the "spatial" vielbein elements $e^i,\;i=1,2,3$, the forms appearing in this equation can not have more than two spatial indices, so in $E_1$ we can find only two $R^{ia}$ forms \textbf{or} one $R^{ia}$ form and one $R^{0i}$ form \textbf{or} one $R^{ij}$ form. It means that if we already have $R^{0i}$ in $E_1$ then we can not have $R^{ij}$ -- this is the reason why we subtract $\alpha_{0i}$ in~(\ref{ij}); we should subtract $\alpha_{0i}$ for $T=1$ only (equations $E_0$ and $E_2$ can contain both $R^{0i}$ and $R^{ij}$ simultaneously), so we use $\delta_{1T}$; in order to avoid negative values of $N_{ij}$ we should subtract $\alpha_{0i}$ only when the number of forms in the equation is more than (or equals to) 2 ($k\geqslant2$) -- we reach it by using the $H(k-2)$ factor. When $k=1$ we have only one form in~(\ref{ET}), so if we already have the form $R^{0i}$ or $R^{0a}$ then we can not have $R^{ij}$, so we subtract the sum $\alpha_{0a}+\alpha_{0i}$; in order to this subtraction works for $k=1$ only we use $\delta_{1k}$. Due to the fact that there are only 3 "spatial" vielbein elements ($e^1,e^2,e^3$) any of the equations $E_0,E_1,E_2$ can not contain more than 3 spatial indices and as a consequence more than 1 form $R^{ij}$, so we subtract the terms $\bgsb{\delta_{1k}(\alpha_{0a}+\alpha_{0i})}$ and $\bgsb{\alpha_{0i}\delta_{1T}H(k-2)}$ from 1.

Explanation for the origin of the formula~(\ref{ia}) are exactly analogous to those above, but much more cumbersome, so we omit it. Finally, replacing $\dot{a}/a$ by $H$ and $\ddot{a}/a$ by $\dot{H}+H^2$ in~(\ref{ET}), we obtain
\eq{\begin{split}
       &E_0\equiv c_0\frac{(D+3)!}{6D!}+c_1\left(\frac{H b'(D+1)!}{b (D-1)!}+\frac{H^2(D+1)!}{D!}+\frac{(\gamma_{D}+b'^2)(D+1)!}{6b^2(D-2)!}\right)+c_2\Biggl(\frac{(\gamma_{D}+b'^2)^2(D-1)!}{6b^4(D-4)!} \\
         & \hspace{1.1cm}+\frac{2H^2(\gamma_{D}+b'^2)(D-1)!}{b^2(D-2)!}+
      \frac{4H^3b'}{b}+\frac{4H^2b'^2(D-1)!}{b^2(D-2)!}+\frac{2Hb'(\gamma_{D}+b'^2)(D-1)!}{b^3(D-3)!}\Biggr) \\
         & +c_3\Biggl(\frac{(\gamma_{D}+b'^2)^3(D-3)!}{6b^6(D-6)!}+\frac{3H^2(\gamma_{D}+b'^2)^2(D-3)!}{b^4(D-4)!}+
      \frac{3Hb'(\gamma_{D}+b'^2)^2(D-3)!}{b^5(D-5)!}+\frac{8H^3b'^3}{b^3} \\
         &\hspace{1.1cm} \frac{12H^2b'^2(\gamma_{D}+b'^2)(D-3)!}{b^4(D-4)!}+\frac{12H^3b'(\gamma_{D}+b'^2)}{b^3}\Biggr)=0
    \end{split}
\label{E0}}
\eq{\begin{split}
       &E_1\equiv c_0\frac{(D+3)!}{2D!}+\\
       &+c_1\left(\frac{2H b'(D+1)!}{b (D-1)!}+\frac{H^2(D+1)!}{D!}+\frac{(\gamma_{D}+b'^2)(D+1)!}{2b^2(D-2)!}+\frac{b''(D+1)!}{b(D-1)!}+\frac{2(H'+H^2)(D+1)!}{D!}\right)+ \\
         &+c_2\Biggl(\frac{(\gamma_{D}+b'^2)^2(D-1)!}{2b^4(D-4)!}+
      \frac{8b''b'H(D-1)!}{b^2(D-2)!}+\frac{4(\gamma_{D}+b'^2)(H'+H^2)(D-1)!}{b^2(D-2)!}+\\
         &\hspace{1.1cm} \frac{4H^2b''}{b}+\frac{4Hb'(\gamma_{D}+b'^2)(D-1)!}{b^3(D-3)!}+\frac{4H^2b'^2(D-1)!}{b^2(D-2)!}+\\
         &\hspace{1.1cm} \frac{2H^2(\gamma_{D}+b'^2)(D-1)!}{b^2(D-2)!}+\frac{8(H'+H^2)Hb'}{b}+\frac{2b''(\gamma_{D}+b'^2)(D-1)!}{b^3(D-3)!}\Biggr)+ \\
         & +c_3\Biggl(\frac{(\gamma_{D}+b'^2)^3(D-3)!}{2b^6(D-6)!}+\frac{3H^2(\gamma_{D}+b'^2)^2(D-3)!}{b^4(D-4)!}+
      \frac{6Hb'(\gamma_{D}+b'^2)^2(D-3)!}{b^5(D-5)!}+\frac{24b''H^2b'^2}{b^3}+ \\
         &\hspace{1.1cm}\frac{12H^2b'^2(\gamma_{D}+b'^2)(D-3)!}{b^4(D-4)!}+\frac{24(H'+H^2)Hb'(\gamma_{D}+b'^2)}{b^3}+\frac{12b''H^2(\gamma_{D}+b'^2)}{b^3}+ \\ &\hspace{1.1cm}\frac{3b''(\gamma_{D}+b'^2)^2(D-3)!}{b^5(D-5)!}+\frac{6(H'+H^2)(\gamma_{D}+b'^2)^2(D-3)!}{b^4(D-4)!}+
         \frac{24b''b'H(\gamma_{D}+b'^2)(D-3)!}{b^4(D-4)!}\Biggr)=0
    \end{split}
\label{E1}}
\eq{\begin{split}
       &E_2\equiv c_0\frac{(D+3)!}{6(D-1)!}+\\
       &+c_1\left(\frac{H b'(D+1)!}{b (D-2)!}+\frac{H^2(D+1)!}{(D-1)!}+\frac{(\gamma_{D}+b'^2)(D+1)!}{6b^2(D-3)!}+\frac{b''(D+1)!}{3b(D-2)!}+\frac{(H'+H^2)(D+1)!}{(D-1)!}\right)+ \\
         &+c_2\Biggl(\frac{(\gamma_{D}+b'^2)^2(D-1)!}{6b^4(D-5)!}+\frac{2H^2(\gamma_{D}+b'^2)(D-1)!}{b^2(D-3)!}+
      \frac{4b''b'H(D-1)!}{b^2(D-3)!}+\frac{4H^3b'(D-1)!}{b(D-2)!}+\\
         & \hspace{1.1cm}\frac{4H^2b''(D-1)!}{b(D-2)!}+\frac{2Hb'(\gamma_{D}+b'^2)(D-1)!}{b^3(D-4)!}+\frac{4H^2b'^2(D-1)!}{b^2(D-3)!}+
         \frac{8(H'+H^2)Hb'(D-1)!}{b(D-2)!}+ \\
         &\hspace{1.1cm}\frac{2b''(\gamma_{D}+b'^2)(D-1)!}{3b^3(D-4)!}+\frac{2(H'+H^2)(\gamma_{D}+b'^2)(D-1)!}{b^2(D-3)!}+4H^2(H'+H^2)\Biggr)\\
         & +c_3\Biggl(\frac{(\gamma_{D}+b'^2)^3(D-3)!}{6b^6(D-7)!}+\frac{3H^2(\gamma_{D}+b'^2)^2(D-3)!}{b^4(D-5)!}+
              \frac{3Hb'(\gamma_{D}+b'^2)^2(D-3)!}{b^5(D-6)!}+ \\
         &\hspace{1.1cm}\frac{12H^2b'^2(\gamma_{D}+b'^2)(D-3)!}{b^4(D-5)!}+\frac{24(H'+H^2)Hb'(\gamma_{D}+b'^2)(D-3)!}{b^3(D-4)!}+
         \frac{24b''H^2b'^2(D-3)!}{b^3(D-4)!}+ \\
         & \hspace{1.1cm}\frac{12b''H^2(\gamma_{D}+b'^2)(D-3)!}{b^3(D-4)!}+\frac{12H^2(H'+H^2)(\gamma_{D}+b'^2)}{b^2}+\frac{24H^2b'^2(H'+H^2)}{b^2}+\\
         &\hspace{1.1cm}\frac{12H^3b'(\gamma_{D}+b'^2)(D-3)!}{b^3(D-4)!}+\frac{8H^3b'^3(D-3)!}{b^3(D-4)!}+\frac{24b''b'H^3}{b^2}\\
         & \hspace{1.1cm}\frac{b''(\gamma_{D}+b'^2)^2(D-3)!}{b^5(D-6)!}+\frac{3(H'+H^2)(\gamma_{D}+b'^2)^2(D-3)!}{b^4(D-5)!}+
         \frac{12b''b'H(\gamma_{D}+b'^2)(D-3)!}{b^4(D-5)!}\Biggr)=0
    \end{split}
\label{E2}}

\section{Numerical analysis\label{num-analysis}}
As it was mentioned above, in the Einstein-Gauss-Bonnet cosmological model the dynamical compactification scenario is realized only in the case when a maximally symmetric solution does not exist. Adding a cubic (i.e. third order in the curvature) term to the Lovelock action leads to a qualitatively different pattern: in this case a maximally symmetric solution co-exists with solutions providing compactification regime. Namely, numerical calculations show that for a given set of coupling constants we get isotropization or compactification regime depending on initial conditions we choose (see Fig.~\ref{graph-max-sym-and-compact} below).

Generally, compactification regime implies that
\eq{b(t)\underset{t\rightarrow\infty}{\longrightarrow}\const\ne0}
On the Fig.~\ref{graph-max-sym-and-compact}a one can see that $\frac{b'(t)}{b(t)}\rightarrow0\Rightarrow b(t)\rightarrow\const$ (Fig.~\ref{suppl-1}) and $H(t)\rightarrow\const$. Any physically realistic regime implies that $H(t)\rightarrow0$ asymptotically. Indeed, even if the observed
cosmological constant in our Universe has a fundamental nature and is not induced by, say, a scalar field,
this value is very small in fundamental units. The requirement $H(t)\rightarrow0$ imposes restrictions on coupling constants and additional restriction on minimal possible number of extra dimensions. Substituting $b''=b'=H'=H=0,\;b=\const\equiv b_{asym}$ and $\gamma_D=-1$ into constraint~(\ref{E0}) and equations of motion~(\ref{E1})-(\ref{E2}), we get equations which we call \emph{asymptotic} in what follows:

\begin{figure}[!t]
\begin{minipage}[h]{.49\linewidth}
\center{\includegraphics[width=.65\linewidth]{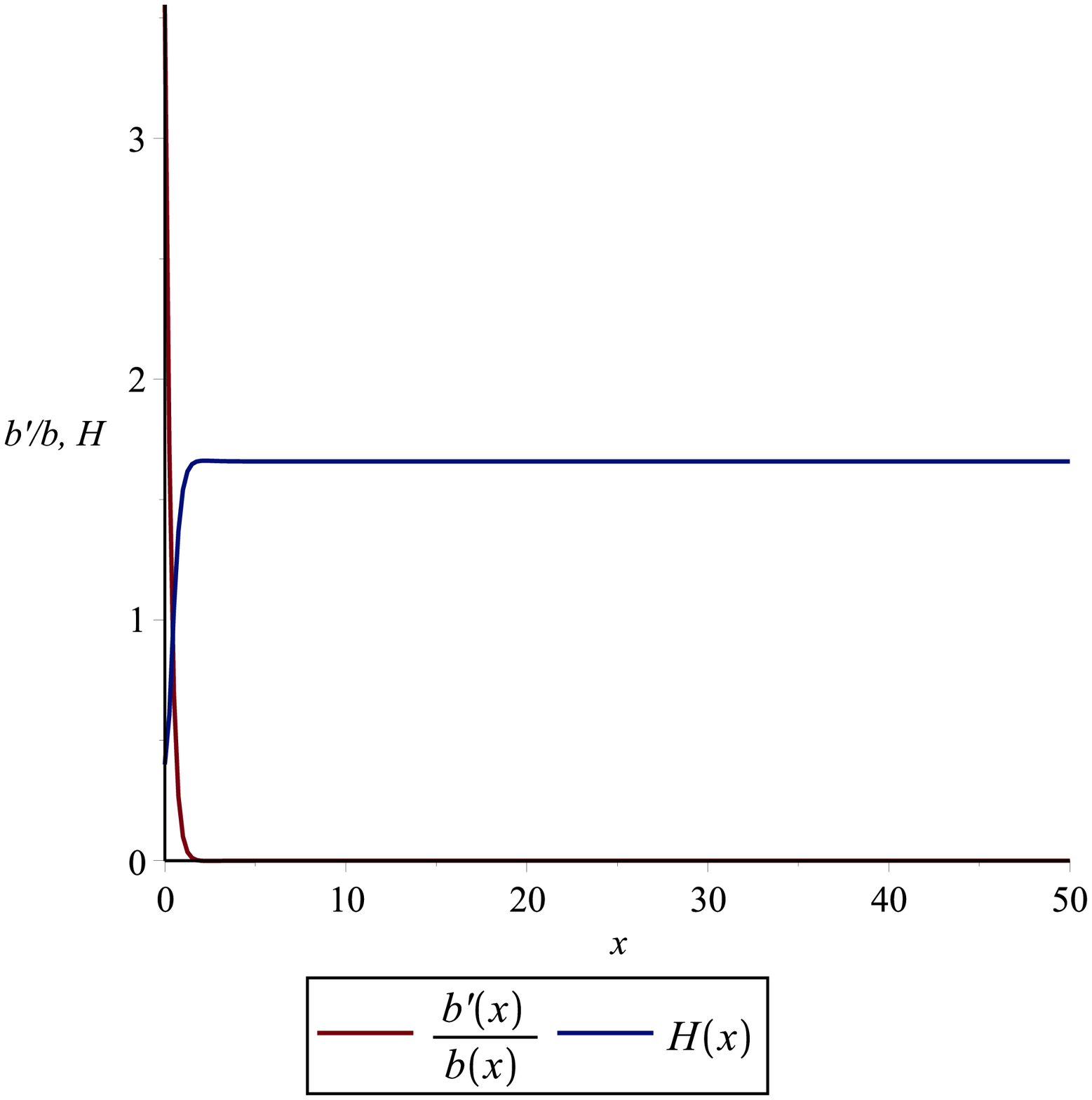} \\ a)}
\end{minipage}
\begin{minipage}[h]{.49\linewidth}
\center{\includegraphics[width=.65\linewidth]{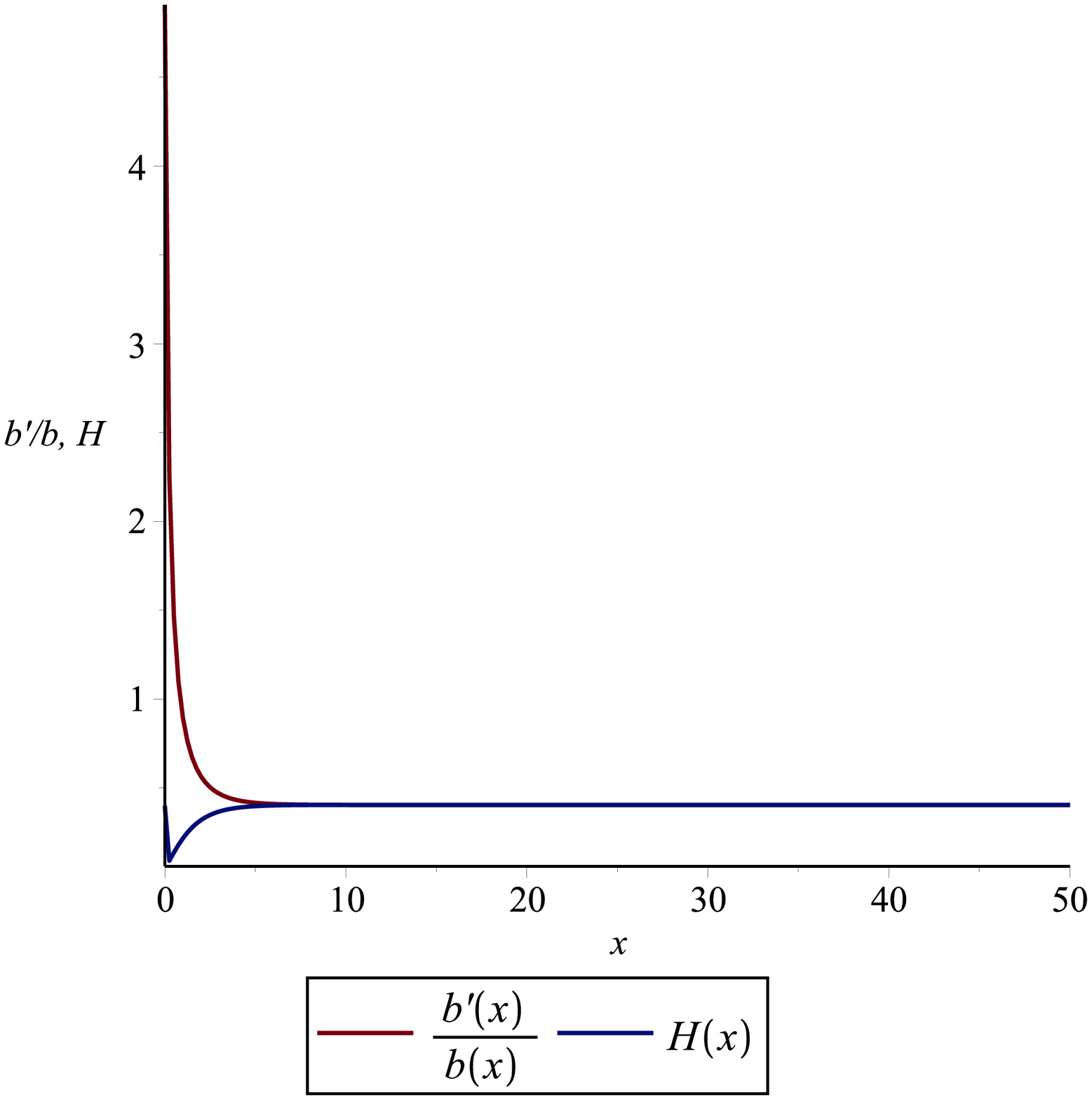} \\ b)}
\end{minipage}
\caption{\footnotesize Compactification and isotropization regimes. Number of extra dimensions $D=7$, coupling constants: $c_0=0.2,\;c_1=-1.1,\;c_2=-0.6,\;c_3=-0.8$, initial conditions: $b_0=0.2,\;H_0=0.4$.  a) For $b'_0=0.7110368731$ we obtain compactification regime. b) For $b'_0=0.9816558121$ we obtain maximally symmetric solution. On these figures $x$ stands for time.}
\label{graph-max-sym-and-compact}
\end{figure}

\begin{multline}\label{asym-1}
  c_0(D+1)(D+2)(D+3)-\frac{c_1(D-1)D(D+1)}{b_{asym}^2}+ \\
  +\frac{c_2(D-3)(D-2)(D-1)}{b_{asym}^4}-\frac{c_3(D-5)(D-4)(D-3)}{b_{asym}^6}=0
\end{multline}
\begin{multline}\label{asym-2}
  c_0D(D+1)(D+2)(D+3)-\frac{c_1(D-2)(D-1)D(D+1)}{b_{asym}^2}+ \\
  +\frac{c_2(D-4)(D-3)(D-2)(D-1)}{b_{asym}^4}-\frac{c_3(D-6)(D-5)(D-4)(D-3)}{b_{asym}^6}=0
\end{multline}
\begin{wrapfigure}{r}{170pt}
   \includegraphics[width=150pt,height=150pt,keepaspectratio]{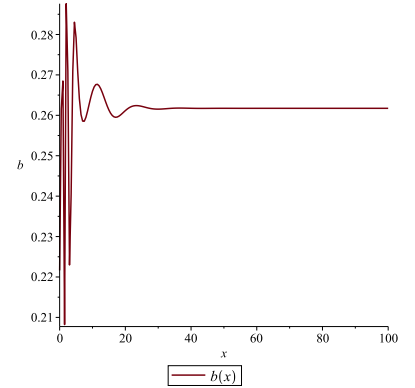}
   \caption{\footnotesize{The supplement to the Fig.~\ref{graph-max-sym-and-compact}. In the compactification regime the scale factor $b$ tend to a non-zero constant asymptotically; $x$ stands for time.}}
\label{suppl-1}
\end{wrapfigure}
Here we took advantage of the fact that the constraint coincides exactly with one of the dynamical equations after the substitution, so we have only two asymptotic equations. These equations are polynomials of degree six with respect to the asymptotic value of the scale factor $b_{asym}$, so the solution for $b_{asym}$ is complicate enough. Taking this into account we solve equations~(\ref{asym-1})-(\ref{asym-2}) for two of the four coupling constants and express them as functions of two other coupling constants, $b_{asym}$ and the number of extra dimensions $D$.

We are looking for compactification regimes which co-exist with the maximally symmetric solution. Maximally symmetric solutions are defined by the equation
\eq{c_3H^6+c_2H^4+c_1H^2+c_0=0 \label{max-sym}}
This equation can be obtained by substituting $b'=bH,\;b''=b(H'+H^2)$ into any of equations~(\ref{E0})-(\ref{E2}). Equation~(\ref{max-sym}) has at least one real root necessarily if coupling constants $c_0$ and $c_3$ have different signs. This fact means that it is convenient to choose any values of $c_0$ and $c_3$ such that $c_0c_3<0$ and then solve equations~(\ref{asym-1})-(\ref{asym-2}) with respect to $c_1$ and $c_2$.

\begin{figure}[!t]
\begin{minipage}[h]{.49\linewidth}
\center{\includegraphics[width=.7\linewidth]{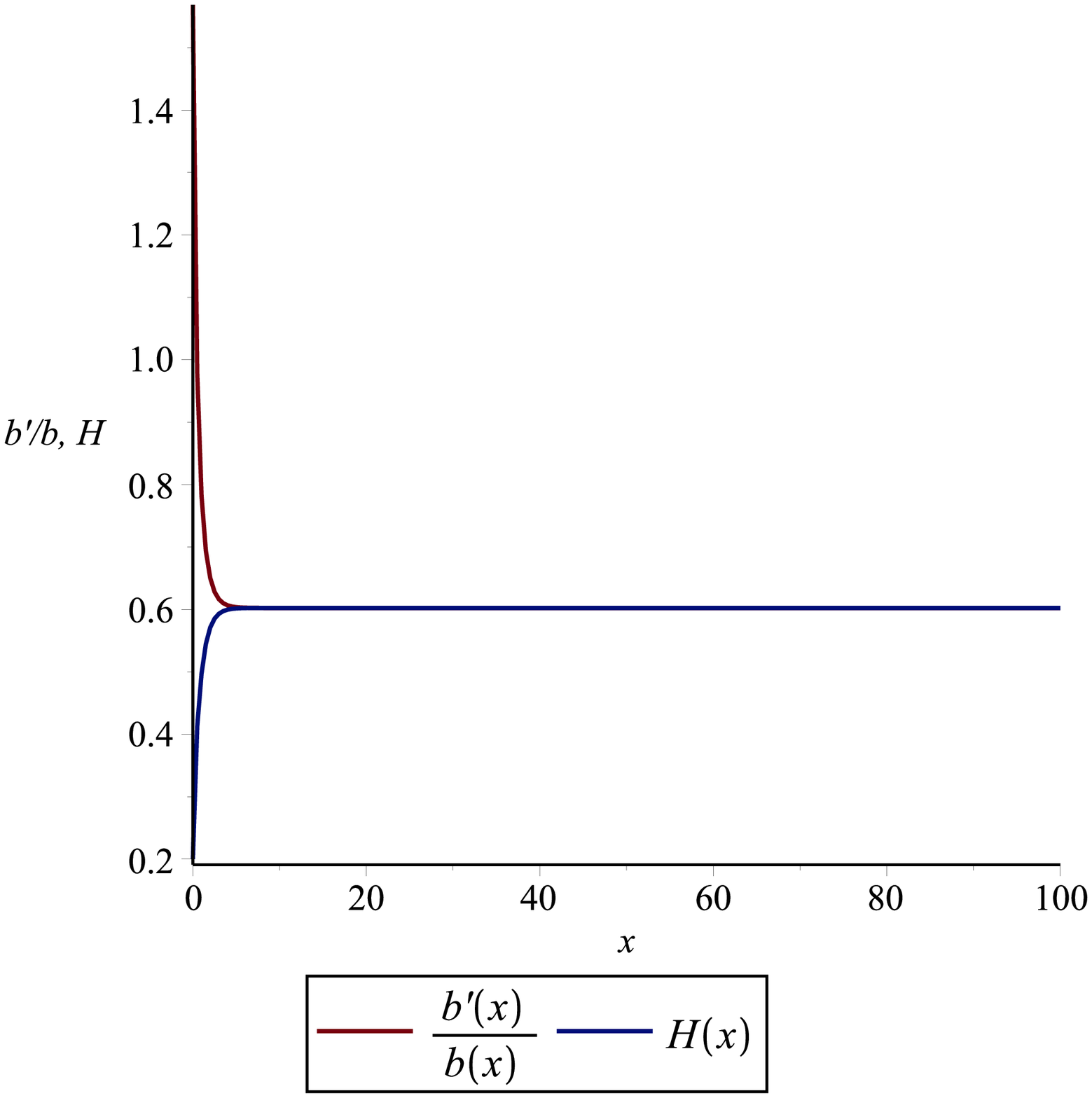} \\ a)}
\end{minipage}
\begin{minipage}[h]{.49\linewidth}
\center{\includegraphics[width=.7\linewidth]{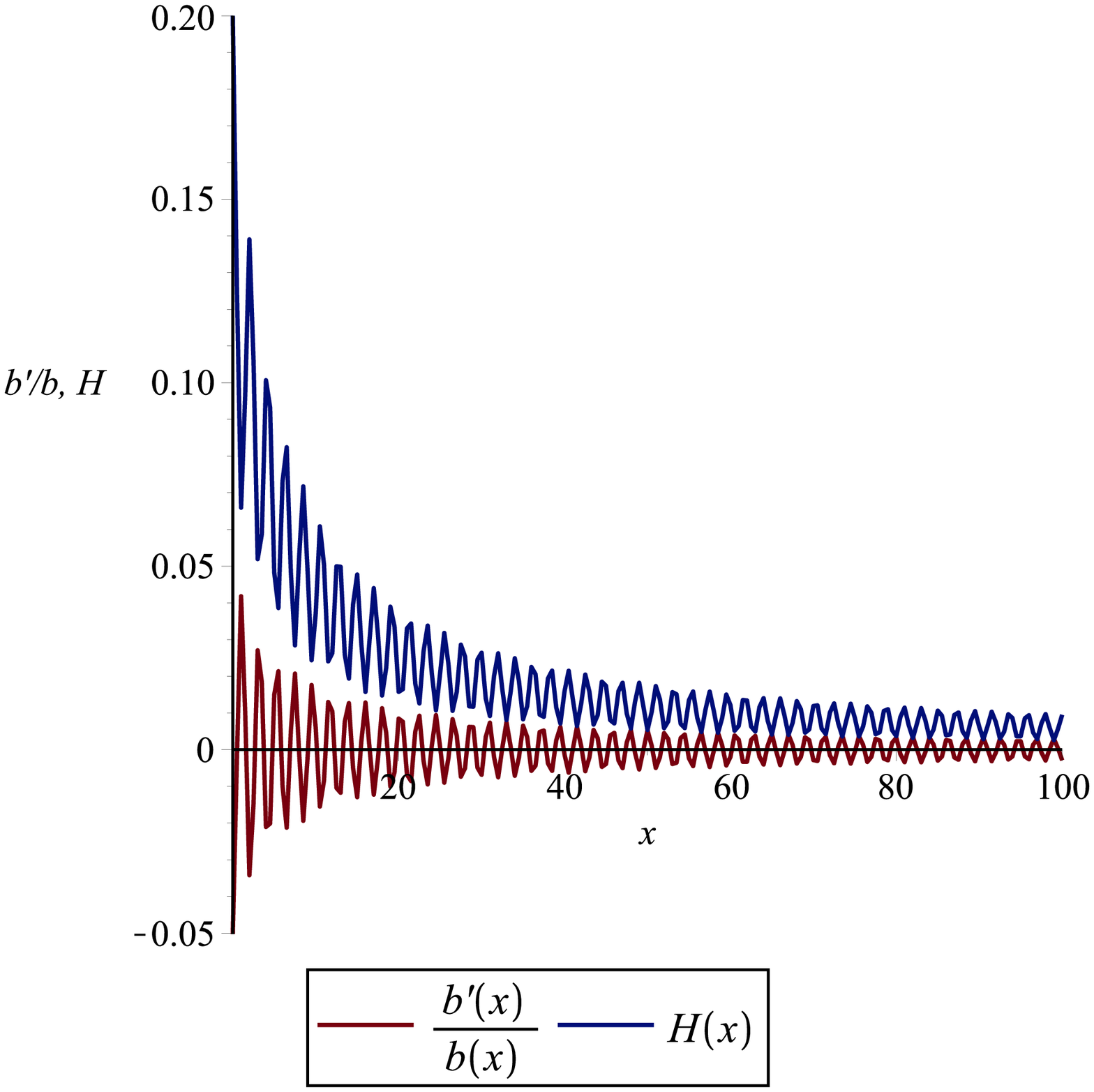} \\ b)}
\end{minipage}
\caption{\footnotesize Compactification and isotropization regimes.  a) Maximally symmetric solution ($b'_0=1.098222529$). b) Compactification ($b'_0=-0.03519064593$). On these figures $x$ stands for time. Number of extra dimensions $D=10$. For the case $D=7$ we have the same (qualitatively) pattern.}
\label{graph-max-sym-and-osc}
\end{figure}

General solution of equations~(\ref{asym-1})-(\ref{asym-2}) with respect to $c_1$ and $c_2$ is cumbersome enough, so we confine ourselves to write down $c_1$ and $c_2$ for particular $D$. It is easy to see from~(\ref{asym-1})-(\ref{asym-2}) that the asymptotic condition $H(t)\rightarrow0$ puts a restriction on the minimal number of extra dimensions: asymptotic regime $H(t)\rightarrow0$ can exist only in models which have at least 6 extra dimensions, otherwise all contributions from 3-d Lovelock term vanish. We consider the cases $D=7$ and $D=10$ and obtain
\eq{c_1=\frac{60c_0b_{asym}^6+c_3}{14b_{asym}^4},\quad c_2=\frac{30c_0b_{asym}^6+2c_3}{5b_{asym}^2}\qquad (D=7)\label{c1c2-7}}
\eq{c_1=\frac{4450c_0b_{asym}^6+147c_3}{1135b_{asym}^4},\; c_2=\frac{2926c_0b_{asym}^6+457c_3}{681b_{asym}^2}\;\; (D=10)\label{c1c2-10}}
\begin{wrapfigure}{r}{170pt}
   \includegraphics[width=150pt,height=150pt,keepaspectratio]{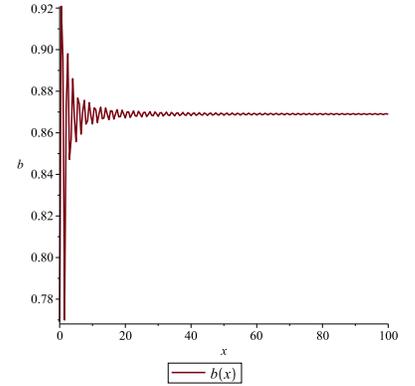}
   \caption{\footnotesize{The supplement to the Fig.~\ref{graph-max-sym-and-osc}. In the compactification regime the scale factor $b$ tend to a non-zero constant asymptotically; $x$ stands for time.}}
   \label{suppl-2}
\end{wrapfigure}
Numerical calculations were performed as follows: we randomly specify values for the couplings $c_0,c_3$ such that $c_0c_3<0$ and the asymptotic value $b_{asym}$ for the scale factor $b(t)$; then we evaluate $c_1, c_2$ from~(\ref{c1c2-7})-(\ref{c1c2-10}); the initial value $H_0$ runs from $0$ to $1$ with a small step, the initial value $b_0$ runs from $(b_{asym}-5)$ to $(b_{asym}+5)$ and the initial value $b'_0$ is evaluated from the constraint~(\ref{E0}). Equation for $b'_0$ is a polynomial of degree six; this polynomial has up to six real roots; numerical calculations show that the minimal of these roots always corresponds to a singular solution, the maximal of them always leads to an isotropic solution (if it exists); the other roots give  singular or/and  compactification solutions. This "distribution"\, is observed both in the case $H(t)\underset{t\rightarrow\infty}{\longrightarrow}0$ and in the case $H(t)\underset{t\rightarrow\infty}{\longrightarrow}\const\ne0$. Thus for the same set of couplings and the same initial values $b_0, H_0$ there exist several regimes: isotropization (maximally symmetric solution), compactification (with oscillatory approach to asymptotic state $b(t)\underset{t\rightarrow\infty}{\longrightarrow}b_{asym},\;H(t)\underset{t\rightarrow\infty}{\longrightarrow}0$) and singularity. The only different feature of the
$H(t)\underset{t\rightarrow\infty}{\longrightarrow}\const\ne0$ general case is the absence of oscillations,
which is natural due to large friction caused by non-zero effective $\Lambda$-term.
Figures~\ref{graph-max-sym-and-osc},\ref{suppl-2} illustrate examples of isotropic solution and compactification; here we specify $c_3=-1.941854169, c_0=0.4491854663, b_0=0.7, H_0=0.2$; from~(\ref{c1c2-10}) we obtain $c_1=-0.1845298663, c_2=-2.196048167$; we find $b'_0$ from the constraint and get four roots: $-1.230556128, -0.03519064593, 0.5896272999, 1.098222529$; the first of them gives singular solution, the next two give compactification regimes and the last one leads to maximally symmetric solution. Note that generally not all the roots with intermediate values correspond to compactification -- some of them can lead to singular solutions (without any regularity).

\begin{figure}[!h]
\begin{minipage}[h]{.49\linewidth}
\center{\includegraphics[width=.7\linewidth]{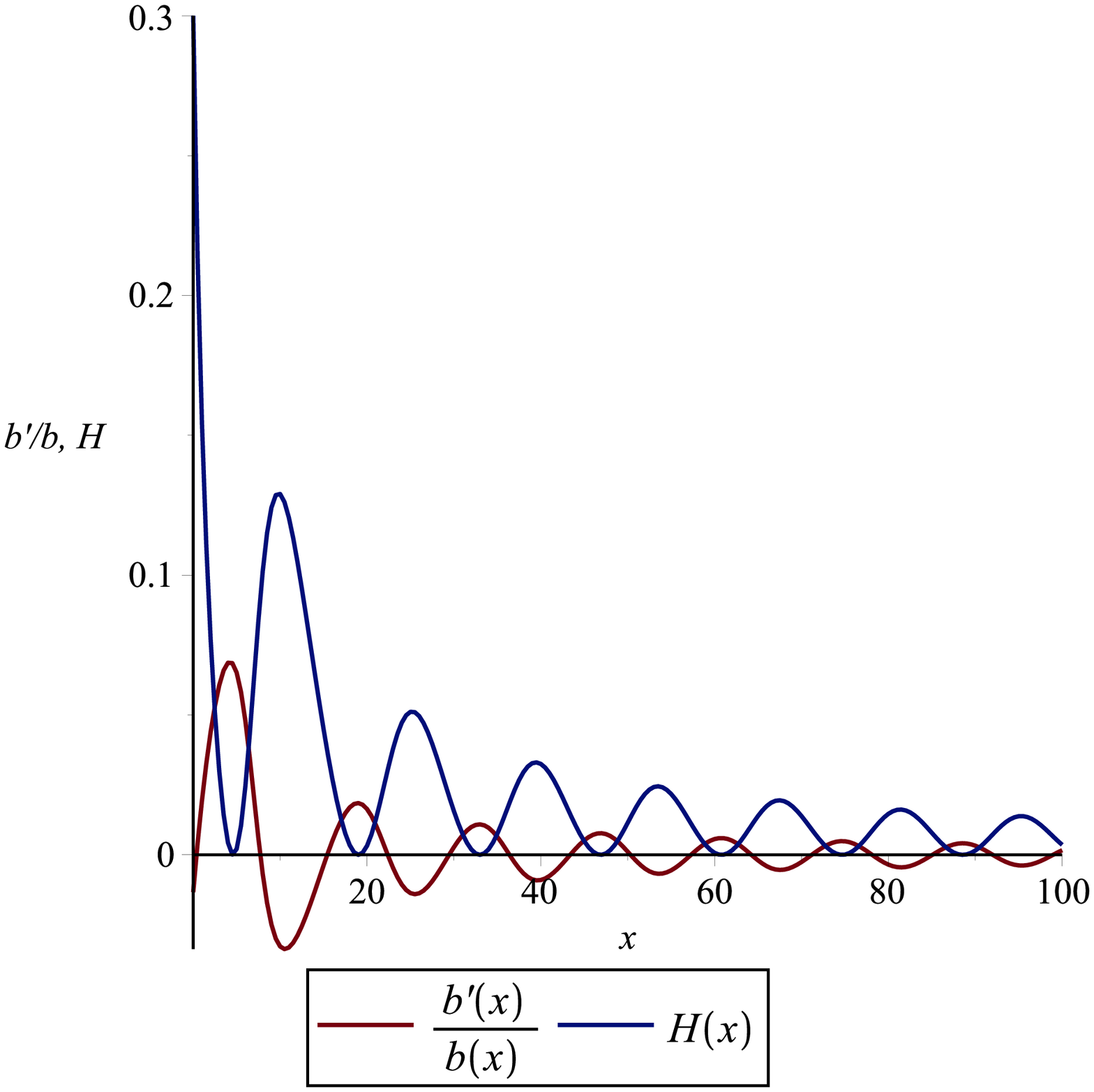} \\ a)}
\end{minipage}
\hfill
\begin{minipage}[h]{.49\linewidth}
\center{\includegraphics[width=.7\linewidth]{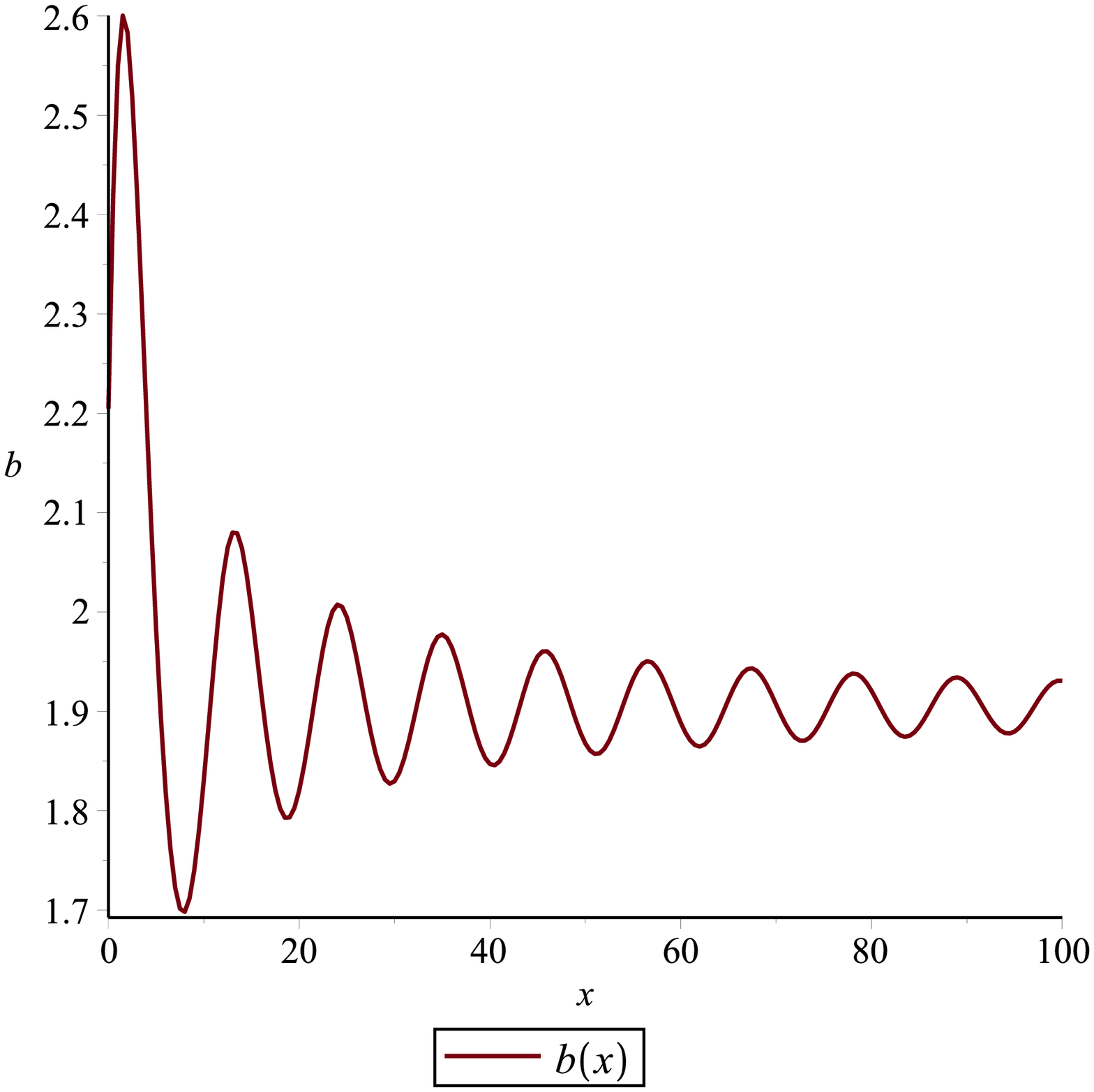} \\ b)}
\end{minipage}
\caption{\footnotesize Compactification regime: number of extra dimensions $D=4$, coupling constants: $c_0=-0.01171782135,\;c_1=-0.2221167296,\;c_2=-3.007375328,\;c_3=-7.54036876$, initial conditions: $b_0=1.345576887,\;b'_0=-0.01797725639,\;H_0=0.3$; $x$ stands for time. Fig. a) shows the behaviour of Hubble parameters; fig. b) demonstrate that the scale factor $b$ tend to a non-zero constant asymptotically.}
\label{D=4}
\end{figure}

Dynamical equations have several summands generated by the cubic Lovelock term which are kept even for $D<6$. For example, for $D=4$ we have
\eq{\begin{split}
      E_0= 35c_0&+c_1\lrp{5H^2+\frac{20Hb'}{b}+\frac{10(-1+b'^2)}{b^2}} \\
        & +c_2\lrp{\frac{4H^3b'}{b}+\frac{12H^2b'^2}{b^2}+\frac{(-1+b'^2)^2}{b^4}+\frac{6H^2(-1+b'^2)}{b^2}+\frac{12Hb'(-1+b'^2)}{b^3}} \\
        & +c_3\lrp{\frac{8H^3b'^3}{b^3}+\frac{3H^2(-1+b'^2)^2}{b^4}+\frac{12H^3b'(-1+b'^2)}{b^3}+\frac{12H^2b'^2(-1+b'^2)}{b^4}}
    \end{split}
}
\eq{\begin{split}
      E_1= 105c_0&+c_1\lrp{15H^2+10H'+\frac{40Hb'}{b}+\frac{30(-1+b'^2)}{b^2}+\frac{20b''}{b}} \\
        & +c_2\Biggl(\frac{12b''(-1+b'^2)}{b^3}+\frac{24b'b''H}{b^2}+\frac{6H^2(-1+b'^2)}{b^2}+\frac{12(H'+H^2)(-1+b'^2)}{b^2}+ \\
        &\hspace{1cm}+\frac{8Hb'(H'+H^2)}{b}+\frac{12H^2b'^2}{b^2}+\frac{3(-1+b'^2)^2}{b^4}+\frac{4H^2b''}{b}+\frac{24Hb'(-1+b'^2)}{b^3}\Biggr) \\
        & +c_3\Biggl(\frac{3H^2(-1+b'^2)^2}{b^4}+\frac{24H^2b''b'^2}{b^3}+\frac{24Hb'b''(-1+b'^2)}{b^4}+\frac{12H^2b'^2(-1+b'^2)}{b^4}+ \\
        &\hspace{1cm}+\frac{6(H'+H^2)(-1+b'^2)^2}{b^4}+\frac{24Hb'(H'+H^2)(-1+b'^2)}{b^3}+\frac{12H^2b''(-1+b'^2)}{b^3}
    \end{split}
}
\eq{\begin{split}
      E_2= 140c_0&+c_1\lrp{\frac{20b''}{b}+40H^2+20H'+\frac{20(-1+b'^2)}{b^2}+\frac{60Hb'}{b}} \\
        & +c_2\Biggl(\frac{12(H'+H^2)(-1+b'^2)}{b^2}+\frac{24b'b''H}{b^2}+4H^2(H'+H^2)+\frac{12Hb'(-1+b'^2)}{b^3}+ \\
        &\hspace{1cm}+\frac{12H^2(b''+Hb')}{b}+\frac{12H^2(-1+3b'^2)}{b^2}+\frac{4b''(-1+b'^2)}{b^3}+\frac{24Hb'(H'+H^2)}{b}\Biggr) \\
        & +c_3\Biggl(\frac{24Hb'(H'+H^2)(-1+b'^2)}{b^3}+\frac{8H^3b'^3}{b^3}+\frac{24H^2b''b'^2}{b^3}+\frac{12H^3b'(-1+b'^2)}{b^3} \\
        &\hspace{1cm}+\frac{24H^2b'^2(H'+H^2)}{b^2}+\frac{12H^2b''(-1+b'^2)}{b^3}+\frac{12H^2(H'+H^2)(-1+b'^2)}{b^2}\Biggr)
    \end{split}
}
Summands generated by the cubic Lovelock term do not alter the compactification solution in EGB gravity with  $H(t)\rightarrow0$, because all
these summands vanish at this solution. However, they, in principle, can change the preceding dynamics.
Numerical calculations show (see Fig.~\ref{D=4}) that these summands do not affect the dynamics of compactification regime which have been  studied in EGB model \cite{CGPT}. This is important since
the number of dimensions needed for the compactification scenario with $H(t)\rightarrow0$ is bigger
than the number for which the next Lovelock term can influence the dynamics. The fact that EGB
compactification solution is still a dynamical attractor when 3-d Lovelock term is taken into account
gives us  a hope that compactification scenario of the present paper will be unaffected by 4-th Lovelock term
(which can not be neglected already for $D=6$), though this needs further investigations.

\section{Conclusions and discussion}
The effect of cubic Lovelock term on the dynamic evolution of compactification in cosmology has been studied. It has been found that the addition of this term does
not spoil the existence of a compactification regime with asymptotic constant three dimensional Hubble parameter and stabilized size of the extra dimensions.\\
This result is surprising because in EGB cosmology in order to achieve this scenario the existence of geometric frustration is crucial. For a cubic theory however there
exist always at least one maximally symmetric solution. A new feature found is that for the cubic theory the compactifying  and isotropizing solutions can coexist which
in EGB was impossible. The results found suggest that these results may be extendible to all Lovelock theories which have an odd curvature power as highest term whereas
the results in EGB gravity may extend to all even power Lovelock theories. It will be an object of future research to check if this conjecture holds.

We also consider a particular version of the compactification scenario when the Hubble parameter in
the 3 large dimensions is (almost) zero. This is needed for the realistic scenario since the effective
cosmological constant in our Universe (if exists at fundamental level and not explained by
some scalar field, for example) is very small in natural units. This additional requirement leads
to one additional relation imposed on the coupling constant of the theory in question. We write down
this relation in a parametric form in order to avoid cumbersome expressions. We note that this particular
regime is present if the number of additional dimensions $D$ is bigger than $5$, otherwise all
contributions from 3-d Lovelock term vanish and we go back to EGB regime. Remembering that
analogous regime in EGB gravity exists for $D>3$, we see an hierarchial structure, similar to
known dimension hierarchy  -- while GB and 3-d Lovelock terms are dynamically important,
correspondingly, for number of extra dimensions $D>0$ and $D>2$, they contribute to compactification
solution with vanishing Hubble constant in large dimensions for $D>3$ and $D>5$.
\section{Acknowledgements} A.G. was partially supported by the FONDECYT grant 1150246 and A.T. was partially supports by the RFBR grant 17-02-01008

\end{document}